\documentclass[fleqn,10pt]{wlscirep}
\usepackage[utf8]{inputenc}
\usepackage[T1]{fontenc}
\usepackage{placeins} % FloatBarrier
\title{Dynamical cross-correlations between RR and QT intervals in long-term electrocardiogram recordings}

\author[1]{Jimi Kokkonen}
\author[1]{Jiyeong Kim-Poikkimäki}
\author[1]{Teemu Pukkila}
\author[1,*]{Esa Räsänen}
\affil[1]{Computational Physics Laboratory, Tampere University, P.O. Box 600, FI-33014 Tampere, Finland}
\affil[*]{esa.rasanen@tuni.fi}

 %up to six allowed

\begin{abstract}
Understanding the dynamic relationship between RR and QT intervals is crucial for interpreting electrocardiograms (ECGs) and managing cardiac conditions. We investigated cross-correlation between RR and QT intervals in long-term ECG recordings from 202 healthy subjects using time-lagged cross-correlation analysis to explore how QT intervals correlate with both preceding and succeeding RR intervals across various time lags. Data was preprocessed with the smoothness priors method to remove long-term trends while preserving relevant short-term dynamics.

Our results reveal that, as expected, the highest RR-QT cross-correlations occur when the intervals overlap the most. Notably, we also observed a strong cross-correlation between the present QT interval and the preceding RR interval, with the overall correlations skewed toward negative lags. This suggests that past RR intervals exert a stronger influence on current QT intervals. These findings align with previous research on RR-QT transfer entropy.

We found no significant differences in RR-QT cross-correlations between sexes. However, aging was associated with significantly reduced cross-correlations across all lags. This effect is consistent with existing literature on age-related cardiac changes and provides further insight into the impact of aging on RR-QT coupling. Additionally, we observed a slight positive correlation between heart rate and cross-correlation at all lags, with cross-correlation decreasing as the lag increased. These findings enhance our understanding of the dynamic RR-QT relationship and may inform future studies on cardiac electrophysiology and diseases.
\end{abstract}

\begin{document}

\flushbottom
\maketitle

\thispagestyle{empty}

%\noindent Please note: Abbreviations should be introduced at the first mention in the main text – no abbreviations lists. Suggested structure of main text (not enforced) is provided below.

\label{ch:introduction}
\section{Introduction}

The relationship between the RR and QT intervals is characterized by a dynamic interplay that is critical for maintaining proper cardiac rhythm. As the heart rate (HR) varies, the QT interval adjusts in response to the RR interval, shortening with increased HRs and lengthening with decreased HRs.~\cite{LuoShen2004Acoc} This adaptive mechanism ensures that the ventricles have adequate time for electrical recovery before the onset of the next cardiac cycle, thus maintaining cardiac efficiency and stability. Understanding the intricate relationship between the RR and QT intervals is vital for the interpretation of the ECG and the management of various cardiac conditions. This relationship underscores the heart’s capacity to adapt its electrical cycle to changing physiological demands. It is well known that the QT-RR relationship exhibits substantial intersubject variability and high intrasubject stability.~\cite{Batchvarov2002}

To account for the influence of HR on the QT interval, the corrected QT interval (QTc) is commonly used in clinical practice. Various correction formulas, such as Bazett’s, Fridericia’s, and Framingham’s, have been developed to adjust the QT interval for a standard HR, thereby enhancing its diagnostic utility.~\cite{LuoShen2004Acoc} Recently, the concept of information transfer from RR to QT intervals has been explored,~\cite{Potapov} leading to an alternative QTc method, such as AccuQT~\cite{AccuQT}, which takes into account the history of RR intervals and aims to minimize the information flow from RR to QT intervals. Accurate assessment of QTc is pivotal in identifying abnormalities, such as Long QT Syndrome and Short QT Syndrome, both of which are associated with an increased risk of life-threatening arrhythmias.~\cite{TseGary2017Emol}

The dynamical adaptation of the QT interval following the changes in HR, often referred to as QT lag, involves both fast and slow phases of adjustment. When there is a sudden change in HR, the QT interval does not adjust instantaneously. Instead, it undergoes a rapid initial adaptation, followed by a slower, more gradual adjustment. This biphasic response is influenced by the autonomic nervous system and the intrinsic properties of cardiac cells.~\cite{Tavernier1997} QT/RR hysteresis refers to the delayed response of the QT interval to changes in the RR interval.~\cite{Gravel2017, Malik2008} This phenomenon is particularly evident during activities such as exercise or stress tests, where the HR increases rapidly, causing the QT interval to lag $\tau$ behind the immediate HR changes. To better understand the QT-RR adaptation dynamics, various mathematical models have been developed. For instance, a recent study integrated an electrophysiological model of a human ventricular cardiomyocyte with a $\beta$-adrenergic signaling cascade model, demonstrating that time-varying $\beta$-adrenergic stimulation plays a crucial role in the QT interval’s adaptation to gradually increasing HR during the exercise phase of a stress test.~\cite{Perez2023-nm}

In this work, we employ time-lagged cross-correlation (TLCC) for RR-QT analysis. TLCC is a statistical method used to analyze the dynamic relationships between two time series by measuring the correlation between them at various time lags.~\cite{Ferreira2023} This approach is particularly useful for studying systems where the interaction or influence between variables may change and involve delayed effects, as is the case with RR and QT intervals. To the best of our knowledge, TLCC has not been previously applied to RR-QT dynamics, although similar concepts, such as lag-based and time-based exponential moving average models, have been used to study the QT/RR hysteresis.~\cite{Jacquemet2017} Here, we demonstrate the utility of TLCC in analyzing the cross-correlation of RR and QT intervals at different time lags in healthy subjects, and how these relationships vary with sex, age, and HR.

\FloatBarrier
% Data and preprocessing
\label{ch:data}
\section{Data and preprocessing}

In this paper, we utilize the E-HOL-03-0202-003 data set from the Telemetric and Holter ECG Warehouse (THEW), referred to here as the THEW dataset. This dataset originates from a population of 202 healthy subjects within the Intercity Digital Electrocardiogram Alliance (IDEAL) database.~\cite{Rochester,CoudercTHEW2010,CoudercTHEW2012} Twenty-four-hour recordings were obtained using a three-lead pseudo-orthogonal configuration and a SpaceLab-Burdick digital Holter recorder. QRS complexes and beat annotations were detected using Spacelab-Burdick Vision Premier software. Prior to each recording, subjects rested in a supine position for 20 minutes. The recordings were sampled at 200 Hz with an amplitude resolution of 10 µV\cite{Rochester}. Descriptive statistics for the population, including mean and standard deviation (mean $\pm$ SD), are presented in Table \ref{table:stats}.

Our dataset exhibited artifacts in the QT intervals, primarily due to the challenges in accurately determining the end of T wave~\cite{AccuQT}. To address the missing or abnormal QT values (where QT > 800 ms or < 150 ms), suitable replacement values were inferred using a k-Nearest Neighbors method, implemented with the Python scikit-learn package~\cite{scikit-learn}. QT interval time series were divided into segments of 5000 beats; first, starting from the beginning of the time series (forward), and second, starting from the end (backward), to include the whole time series. In each segment, missing values were imputed using the five nearest neighboring samples, and the forward and backward time series were recombined. The final QT time series was obtained by averaging these two imputed series.

Additionally, when calculating cross-correlations across the entire time series, a detrending procedure, based on smoothness priors~\cite[pp. 27-31]{Genshiro}, was employed. This procedure involved a range of smoothing parameters $\gamma$ from 10 to 500, which correspond to the cutoff frequency of the filter~\cite{TarvainenSmoothnessPriors}.

Detrending is crucial for removing obvious trends in the RR/QT data that might obscure the relevant dynamic correlations between RR and QT intervals. A larger $\gamma$ value filters out more low-frequency components of the signal, while a smaller $\gamma$ retains these components but removes higher-frequency fluctuations. Although detrending reduces data roughness and can minimize noise-induced false correlations, it may also eliminate genuine correlations with underlying phenomena. Thus, the parameter $\gamma$ represents a trade-off between the accuracy of the detrending procedure and the smoothness of the resulting data~\cite[pp. 29]{Genshiro}. The impact of detrending on the RR-QT cross-correlations is analyzed in detail below.

\begin{table}[h!]
\centering
\caption{Descriptive statistics (mean $\pm$ SD) of the population in the THEW dataset.}
\begin{tabular}{cccc}
{} & Female & Male & All \\
 \hline
N      &  100 & 102 & 202 \\
Age (years)   &   40.4 $\pm$ 17.2 &    36.6 $\pm$ 14.1 &   38.5 $\pm$ 15.8 \\
Height (cm) &   162.3 $\pm$ 6.2 &   176.9 $\pm$ 8.9 &  169.7 $\pm$ 10.6 \\
Weight (kg) &  62.3 $\pm$ 13.4 &   77.5 $\pm$ 12.9 &    70.0 $\pm$ 15.2 \\
BMI (kg/m²)    &   23.7 $\pm$ 5.4 &  24.6 $\pm$  3.1 &   24.2 $\pm$ 4.4\\
\end{tabular}
\label{table:stats}
\end{table}

\FloatBarrier
% Theory and methods
\label{ch:methods}
\section{Time-lagged cross-correlation method}

The basic cross-correlation function is a fundamental technique that underpins many advanced analytical methods. It quantifies the relationship between two time series by first computing the {\em cross-covariance function} of the series:\cite{Penny}

\begin{equation}
\sigma_{xy} = \frac{1}{N-1}\sum^{N}_{i=1}(x_i-\mu_x)(y_i-\mu_y),
\end{equation}

where $\sigma_{xy}$ is the cross-covariance, $N$ is the number of samples in each time series, $x_i$ is the $i$th point of the first time series, $y_i$ is the $i$th point of the second time series, and $\mu_x$ and $\mu_y$ are the mean values of the first and second time series, respectively. The removal of the mean values of each time series is done in order to improve results by accentuating the cross-covariance values. 

From the cross-covariance, the {\em cross-correlation function} can be calculated with a normalization procedure:
\begin{equation}
r_{xy} = \frac{\sigma_{xy}}{\sqrt{\sigma_{xx}\sigma_{yy}}},
\end{equation}
where $\sigma_{xx}=\sigma_{x}^2$ and $\sigma_{yy}=\sigma_{y}^2$ are the variances of each signal, calculated similarly to the cross-covariance as follows:

\begin{equation}
\sigma_{x} = \frac{1}{N}\sum^{N}_{i=1}(x_i-\mu_x)^2,
\end{equation}

and

\begin{equation}
\sigma_{y} = \frac{1}{N}\sum^{N}_{i=1}(y_i-\mu_y)^2.
\end{equation}

A higher absolute value of the cross-correlation coefficient $r_{xy}$ indicates a greater similarity between the two time series. Curves that rise and fall simultaneously will exhibit a positive cross-correlation, while cases where one curve rises and the other falls will show a negative cross-correlation, reflecting an inverse relationship.

In addition, one can shift the time series relative to each other by a certain lag $\tau$ and then calculate the cross-correlation between these shifted time series, as follows:
\begin{equation}
\sigma_{xy}(\tau) = \begin{cases}
    \displaystyle \frac{1}{N - \tau}\sum^{N - \tau}_{i=1}(x_{i+\tau}-\mu_x)(y_i-\mu_y), & \tau \geq 0 \\
    \displaystyle \frac{1}{N - |\tau|}\sum^{N - |\tau|}_{i=1}(x_i-\mu_x)(y_{i + |\tau|}-\mu_y), & \tau < 0
\end{cases}
\end{equation}
and
\begin{equation}
r_{xy}(\tau) = \frac{\sigma_{xy}(\tau)}{\sqrt{\sigma_{xx}(\tau)\sigma_{yy}(\tau)}},
\end{equation}
where the variances and means are calculated over the whole time series.

\begin{figure}[t]
\includegraphics[width=\textwidth]{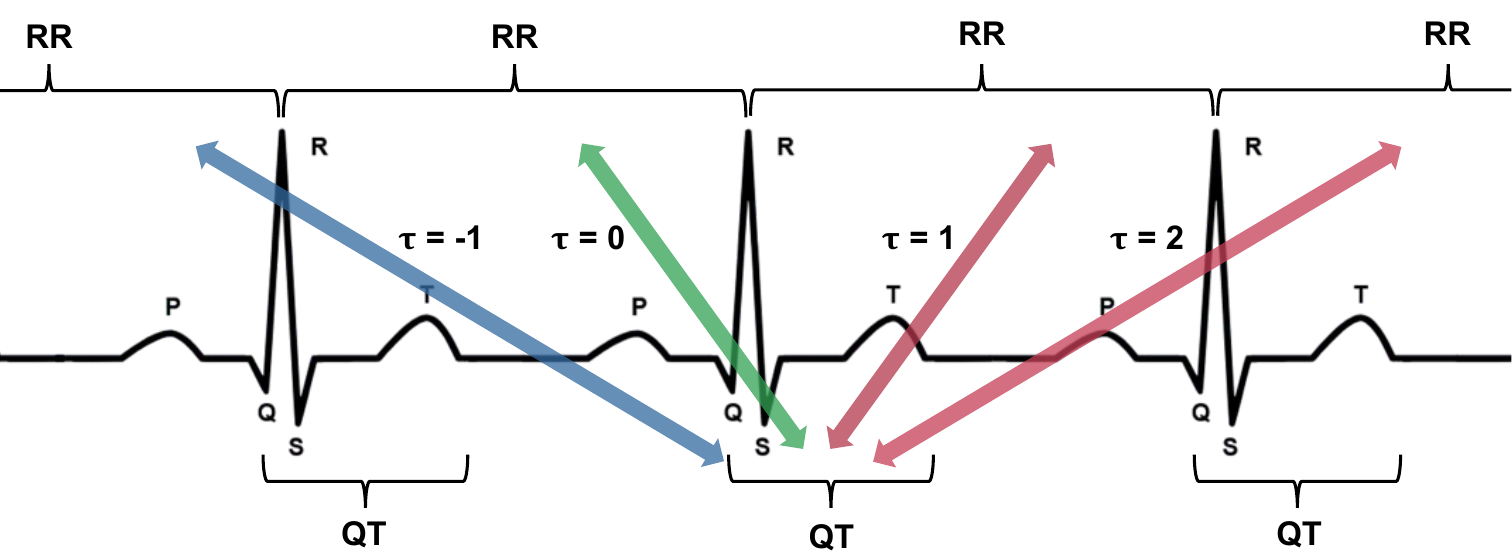}
\caption[Example of lags between RR and QT intervals in an ECG signal.]{Example of lags between RR and QT intervals in an ECG signal. Brackets at the top indicate RR intervals, while those at the bottom indicate QT intervals. Colored arrows indicate the sign of the lag, with negative lags in blue, zero lag $\tau$ in green and positive lags in red.}
\label{fig:lags}
\centering
\end{figure}

Figure \ref{fig:lags} illustrates how different values of lag $\tau$ are applied in the cross-correlation analysis of an ECG signal. It is important to note that a single RR interval overlaps with two different QT intervals: the preceding QT interval, which has a significant amount of overlap, and the succeeding QT interval, which has a smaller amount of overlap. We define $\tau = 0$ to correspond to comparing an RR interval with the succeeding QT interval. This choice aligns with conventional QTc methods, where QT intervals are corrected based on the preceding RR interval. With $\tau = 1$, the QT time series is shifted forward by one interval, and the cross-correlation is calculated between the RR interval and the next QT interval. Cross-correlations between more distant ECG events are calculated using negative lags and lags greater than 1.

\FloatBarrier
% Statistical analysis
\label{ch:statistical_analysis}
\section{Statistical analysis}

Student's t-test and Welch's t-test were employed to determine the statistical significance of the difference in relevant parameters, with a significance level of 5\%. In particular, a dependent sample t-test was used to compare the mean cross-correlation between two smoothness prior detrending parameters for each lag $\tau$ in Sec.~\ref{results:detrending}, and Welch's t-test to compare between two gender groups for each lag $\tau$ in Sec.~\ref{results:gender}. A dependent sample t-test was also used to compare the HR dependence between the lags in Sec.~\ref{results:hr_dependence}. In each analysis, the normality assumption holds under the central limit theorem due to the large sample size. Due to the identical sample sizes in both groups, the test is also robust for uneven variances.~\cite{MarkowskiConditionsForVariance}
\FloatBarrier
% Results
\label{ch:results}
\section{Results and discussion}

\subsection{Whole population and effects of detrending}\label{results:detrending}

%\begin{figure}[t]
%\centering
%\includegraphics[width=0.75\textwidth]{figures/Gamma comparison of RR-QT cross-correlation across entire subjects, gamma1 = 10, gamma2 = 500.pdf}
%\caption{Box plots of RR-QT cross-correlations as a function of lag for all subjects (N = 202) with detrending parameters $\gamma = 10$ and $\gamma = 500$. The box plots mark the first %and third quartiles (the top and bottom of each box), the median (the line inside each box), the minimum and maximum (the whiskers below and above each box) as well as outliers (the %circles). The p-values from Student's independent two-sample t-test are also shown for each pair of $\gamma$ for each lag. The gray horizontal line marks the point where cross-%correlation becomes negative.}
%\label{fig:rr_qt_CCs_10_and_500}
%\centering
%\end{figure}

\begin{figure}[t]
\centering
\includegraphics[width=\textwidth]{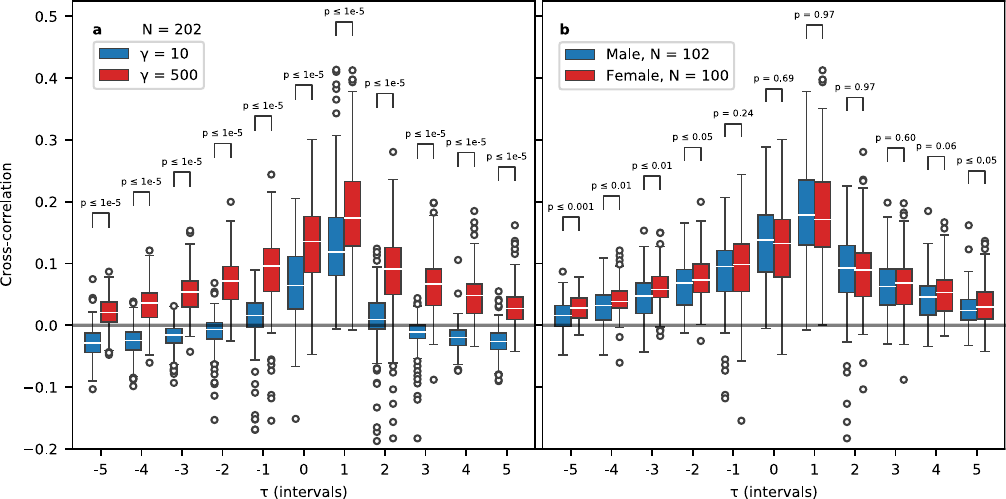}
\caption{(a) Box plots of RR-QT cross-correlations as a function of lag for all subjects (N = 202) with detrending parameters $\gamma = 10$ and $\gamma = 500$. (b) Box plots of RR-QT cross-correlations as a function of lag for men (N = 102) and women (N = 100). The box plots show the first and third quartiles (the top and bottom of each box), the median (the white line inside each box), the minimum and maximum values (the whiskers below and above each box), as well as outliers (the circles). The gray horizontal line indicates the point where cross-correlation becomes negative. Extreme outliers with cross-correlations above 0.5 or below 0.2 have been removed for display purposes. The p-values from Student’s t-test are shown for each pair at each lag, comparing (a) the $\gamma$ values and (b) the sexes.
}
\label{fig:CC_sex_and_gamma_compare}
\centering
\end{figure}

Figure \ref{fig:CC_sex_and_gamma_compare}(a) shows the time-lagged cross-correlations between RR and QT intervals for the complete dataset (N=202) in two smoothness priors detrending parameters ($\gamma=10$ and $500$). We consider lags $\tau=-5,\ldots, 5$, as defined in Fig.~\ref{fig:lags}, i.e., $\tau = 0$ corresponds to the pair of the present QT interval and the previous (partly overlapping) RR interval. %The statistical significance of the difference in the mean cross-correlation between two $\gamma$ values was tested for each lag. Student's paired samples t-test was used ($\alpha=0.05$) as the two samples are related. Normality assumption holds under the central limit theorem due to the large sample size, and due to the identical sample sizes in both groups, the test is also robust for uneven variances.~\cite{MarkowskiConditionsForVariance}

We find that the highest cross-correlation, regardless of the detrending parameter, is achieved with $\tau = 1$. At this lag, the RR and QT intervals exhibit the greatest mutual overlap, as shown in Fig.\ref{fig:lags}, making this result plausible. The second highest cross-correlation is observed with $\tau = 0$. At this lag, there is small overlap between the intervals (QR interval), which is not significant. Therefore, the relatively high correlation between the QT interval and the \textit{preceding} RR interval aligns with previous findings on the information transfer between RR and QT intervals.\cite{Potapov} Overall, the profile of correlations across lags $\tau = -5 \ldots 5$ is skewed towards negative lags, indicating that the history of RR intervals has a greater impact on the QT interval than vice versa. This observation is consistent with the transfer entropy results.~\cite{Potapov}

Figure \ref{fig:CC_sex_and_gamma_compare}(a) illustrates that using high-frequency detrending with a small smoothing parameter $(\gamma = 10)$ generally reduces cross-correlation across all lags. This suggests that short-range trends do influence the cross-correlation. Despite this reduction, the overall pattern of the largest cross-correlations $(\tau = 1, 0, -1, -2)$ remains consistent on average. However, with this level of detrending, other lags do not exhibit detectable cross-correlations. In the following results, we use $\gamma=500$ throughout the paper.

\subsection{Effects of sex and age}\label{results:gender}

% --- SEX ---

%\begin{figure}[t]
%\centering
%\includegraphics[width=0.75\textwidth]{figures/Gender comparison of RR-QT cross-correlation across entire subjects, gamma = 500.pdf}
%\caption{Box plots of RR-QT cross-correlations as a function of lag for men (N=102) and women (N=100). The p-values from Student's two-sample independent t-test are shown for each pair %of sexes at each lag.}
%\label{fig:rr_qt_CCs_genders_500}
%\centering
%\end{figure}

Figure \ref{fig:CC_sex_and_gamma_compare}(b) shows the cross-correlation as a function of lag, with all subjects divided into male and female groups. %Statistical significance between each pair of sexes was tested with Welch's t-test ($\alpha=0.05$) as the two populations are separate with unequal variances and sample sizes. Normality is assumed under the central limit theorem due to large sample size. 
We find that at lags $\tau = -2, \ldots, 4$, sex does not have a statistically significant effect on the cross-correlation. Beyond this range of lags, females exhibit slightly higher cross-correlations. However, no definitive conclusions can be drawn from this trend, even though it is known that sex influences the cardiac cycle and electrocardiogram, with women typically having shorter QRS and longer QT intervals than men.\cite{Rautaharju} Age is not a confounding factor in this analysis, as the age distributions between the two groups are very similar (see Table~\ref{table:stats}).

% --- AGE ---

\begin{figure}[t]
\centering
\includegraphics{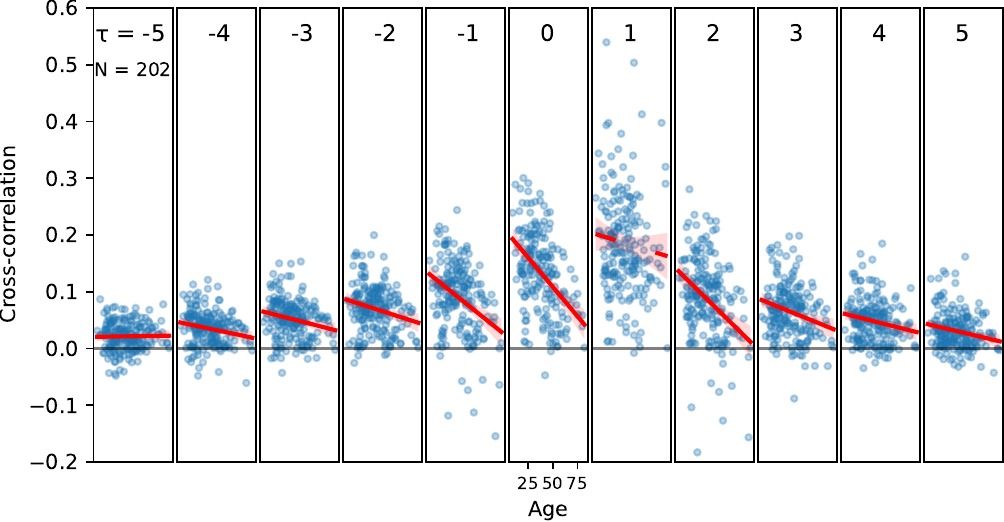}

\caption[Scatter plots of cross-correlation as function of age at lags $\tau$ from 5 to 5.]{Scatter plot of cross-correlation as function of age for each lag from -5 to 5. The red lines indicate linear regression fits on each scatter plot. The 95\% confidence interval is shown in light red.}
\label{fig:cc_vs_age_scatter}
\end{figure}

%\begin{table}
%\centering
%\caption{Statistics for the linear regression of age versus cross-correlation for each lag. Included are the P-value, the $r^2$ value and the mean squared error (MSE) of the residuals.}
%\begin{tabular}{lrrr}
%Lag &       p-value &        r² &  MSE of %residuals \\
%\hline
%-5 & 0.810 & 0.000 & 0.001 \\
%-4 & 0.006 & 0.038 & 0.001 \\
%-3 & 0.004 & 0.042 & 0.001 \\
%-2 & 0.002 & 0.046 & 0.002 \\
%-1 & 0.000 & 0.125 & 0.003 \\
%0 & 0.000 & 0.212 & 0.004 \\
%1 & 0.241 & 0.007 & 0.010 \\
%2 & 0.000 & 0.132 & 0.005 \\
%3 & 0.001 & 0.056 & 0.002 \\
%4 & 0.009 & 0.034 & 0.001 \\
%5 & 0.005 & 0.040 & 0.001 \\
%\end{tabular}
%\label{table:age_scatter_stats}
%\end{table}

\begin{table}
\centering
\caption{Statistics for the linear regression of age versus cross-correlation for each lag. Included are the P-value, the $r^2$ value and the mean squared error (MSE) of the residuals.}
\begin{tabular}{cccc|cccc}
Lag & p-value & r² & MSE of residuals & Lag & p-value & r² & MSE of residuals \\
\hline
-5 & 0.810 & 0.000 & 0.001 & 1 & 0.241 & 0.007 & 0.010\\
-4 & 0.006 & 0.038 & 0.001 & 2 & 0.000 & 0.132 & 0.005 \\
-3 & 0.004 & 0.042 & 0.001 & 3 & 0.001 & 0.056 & 0.002\\
-2 & 0.002 & 0.046 & 0.002 & 4 & 0.009 & 0.034 & 0.001\\
-1 & 0.000 & 0.125 & 0.003 & 5 & 0.005 & 0.040 & 0.001\\
0 & 0.000 & 0.212 & 0.004 &&&&\\
\end{tabular}
\label{table:age_scatter_stats}
\end{table}

Figure~\ref{fig:cc_vs_age_scatter} presents scatter plots of cross-correlation as a function of the subject’s age, separated by different lags of $\tau = -5, \ldots, 5$. An ordinary least squares (OLS) linear regression fit (in red) is applied to each scatter plot, with the $95\%$ confidence interval shown in light red. Notably, the slope of the regression line is negative at all lags (except $\tau=-5$), suggesting that aging reduces the cross-correlation between RR and QT intervals. The reduction in cross-correlation is most pronounced at the smallest lags. For instance, at $\tau = 0$, the cross-correlation decreases on average by about two-thirds as age increases from 25 to 75 years. However, it is important to note that the sample size decreases significantly with age, which complicates the interpretation of these results. Despite this limitation, the effect of age remains substantial.
Table ~\ref{table:age_scatter_stats} displays the p-value, $r^2$ value and mean squared error of the residual for the OLS fit for each lag. Both negative and positive lags have increased $r^2$ values and decreased p-values when approaching lag $\tau = 0$. Lag $\tau = 1$ is the exception showing small $r^2$, a statistically insignificant p-value ($\alpha$ = 0.05), and a high mean squared error of residuals. This is likely explained by the significant overlap between RR and QT intervals at lag $\tau = 0$, as discussed earlier.

The effects of aging on cross-correlations are consistent with previous findings. Baumert et al.\cite{Baumert2013} demonstrated that as individuals age, there are notable changes in cardiac electrophysiology, including QT interval prolongation and increased QT variability. Aging is particularly associated with a decoupling of QT variability from heart rate variability, as evidenced by various methods such as cross-multiscale entropy, information-based similarity index, and joint symbolic dynamics. Therefore, the observed reduction in cross-correlation as a function of age, as shown in Fig.~\ref{fig:cc_vs_age_scatter}, aligns with these earlier studies.

\subsection{Heart-rate dependence}\label{results:hr_dependence}

So far, we investigated RR-QT cross-correlation calculated over 24-hour recordings. While the detrending procedure removes the slow fluctuations in HR, the effects of local HR on cross-correlation are averaged out. To investigate the intrinsic HR dependence of RR-QT cross-correlation, analysis was conducted on 100-beat segments from individual RR and QT time series. The length of the segments was chosen to be long enough that cross-correlation can be computed reliably, avoiding spurious correlations from calculating across excessively short segments, yet short enough that the mean beat rate can be defined reasonably to approximate a local HR dependence as accurately as possible. The 100-beat-long segments within subjects had a median standard deviation of HR ranging from 1.7 bpm to 16.6 bpm, with an overall median of 5 bpm. One subject with a large variation in HR ($>30$ bpm) within short segments was omitted from this analysis. 

\begin{figure}[t]
\centering
\includegraphics[width=\linewidth]{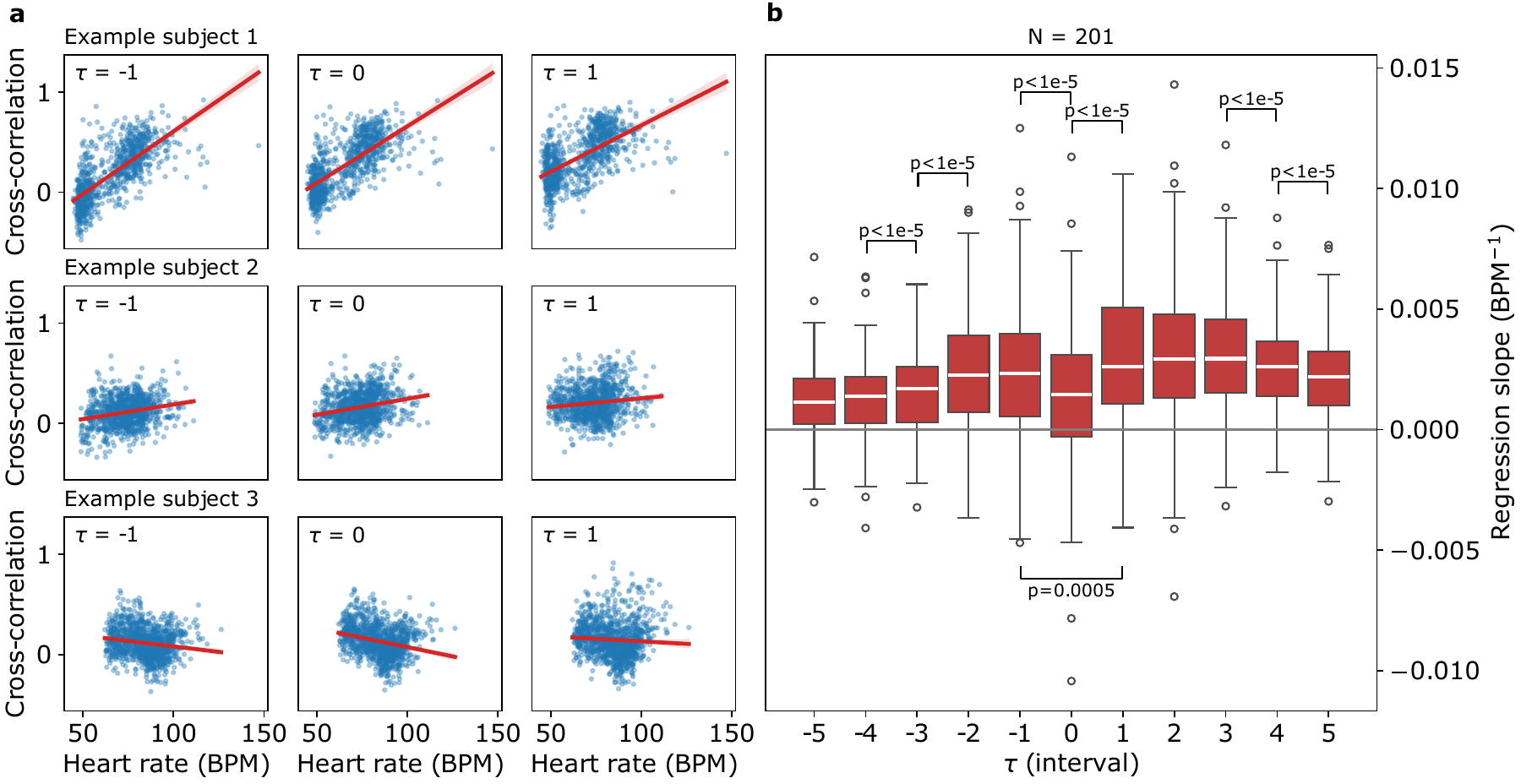}
\caption{(a) Example scatter plots of RR-QT cross-correlations at three lags (-1, 0, 1) versus heart rate (HR) with corresponding regression fits for a subject with positive slopes (top row), a subject with slightly positive slopes (middle row), and a subject with negative slopes (bottom row). (b) Box plot of regression slopes from individual RR-QT cross-correlations calculated in short segments. P-values of a few selected pairs of lags are presented.}
\label{fig:xcorr_hr_box}
\centering
\end{figure}

For each subject, RR-QT cross-correlation and the average HR were calculated for each segment, and their correlation was estimated by the regression slope, as shown in the examples in Fig.~\ref{fig:xcorr_hr_box}(a). The regression slopes at different lags for all subjects are summarized in Fig.~\ref{fig:xcorr_hr_box}(b). Subject-to-subject variations were clearly present in the relationship between RR-QT cross-correlation and average heart rate, as suggested by the examples showing strongly positive, weakly positive, and negative slopes in Fig.~\ref{fig:xcorr_hr_box}(a). However, most subjects exhibit positive regression slopes across all lags, suggesting a positive correlation between cross-correlation and HR. 

The main finding is that the mean slope, indicating the relationship between RR-QT cross-correlation and HR, changes with lags, with several notable characteristics. First, we note that the HR dependence is suppressed at lag $\tau = 0$. It immediately increases when a single lag $\tau$ in either direction is introduced. Because cross-correlation between RR and QT decreases with larger (positive and negative) lags, it follows that the HR dependence diminishes with increasing lags in either direction, due to the positive correlation between RR-QT cross-correlation and HR. We also note that the HR dependence is asymmetric around $\tau = 0$, with positive lags having significantly stronger dependence on HR. This may be due to two different factors acting on the cross-correlation between RR and QT: RR-dominant RR-QT dynamics and the presence of a trend. These two factors affect the regression slopes in Fig.~\ref{fig:xcorr_hr_box}(b) simultaneously. Due to RR hysteresis~\cite{Gravel2017, Malik2008} and RR's dominance over the succeeding QT~\cite{Potapov}, cross-correlation between RR and QT is asymmetric around lag $\tau=1$, with negative side of the lags having higher cross-correlation, as we have seen in Fig.~\ref{fig:CC_sex_and_gamma_compare}. Therefore, for lags $\tau=0,\ -1,\ -2,\ \cdots$, cross-correlation is higher on the lower HR regime, resulting in smaller (positive) regression slopes. On the other hand, while the low HR segments likely represent steady resting states, high HR segments likely contain sudden bursts of activity and thus have relatively more trends. The presence of trends in the high HR segments increases cross-correlation on all lags, resulting in overall larger regression slopes in positive lags.

% --- MEAN CC ---

\begin{figure}[t]
\centering
\includegraphics{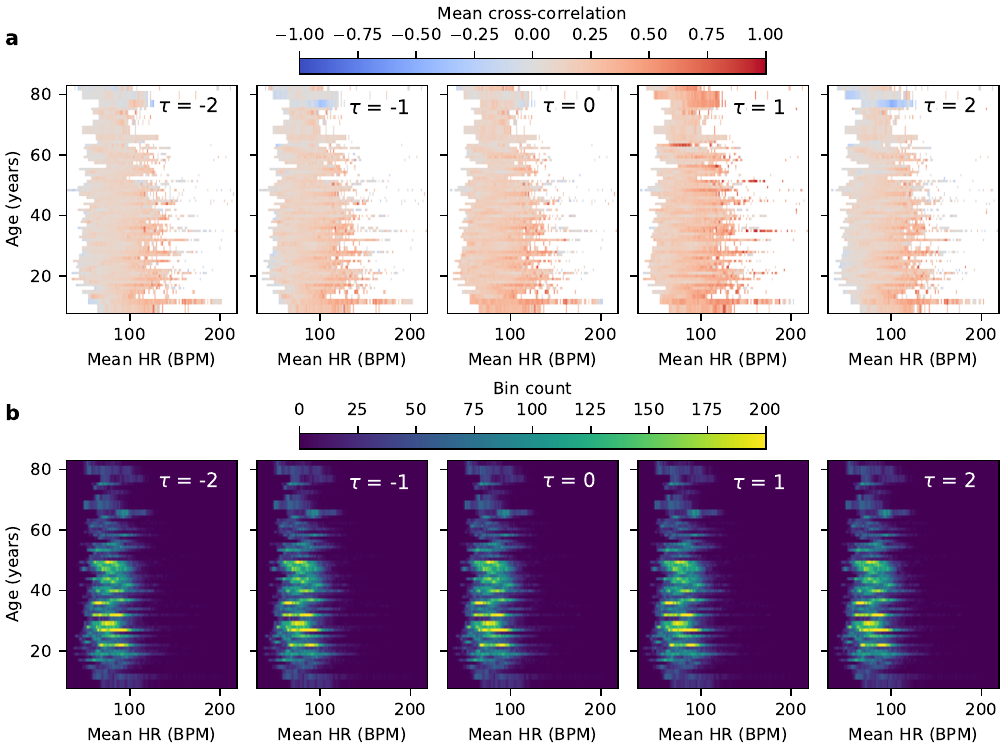}
\caption{Mean cross-correlation as a function of age and mean heart rate in 100 beat segments at lags from -2 to 2. Age and mean heart rate are binned as integer values and the mean cross-correlation is calculated with Fisher $z$-transform to avoid bias~\cite{Silver}. The bottom panel shows count of values in each bin for assessing the statistical properties of the mean cross-correlation.}
\label{fig:age_hr_xcorr}
\end{figure}

In Fig.~\ref{fig:age_hr_xcorr}, the effects of both age and HR on cross-correlation are analyzed. Cross-correlation and mean HR were calculated for 100-beat segments across all subjects. Mean HR values were then grouped into integer bins, and the mean cross-correlation in each bin was plotted as a function of mean HR and the subject’s age. To avoid bias, mean cross-correlation was calculated using Fisher’s $z$ transformation \cite{Silver} \cite{ToivonenArticle}. These heatmap plots are shown in the top panel of Fig.~\ref{fig:age_hr_xcorr} for 100-beat-long segments. Additionally, to assess the impact of value counts in each bin on the results, the count of values in each bin was plotted in the bottom panel.

The results confirm previous findings that lag $\tau=1$ has the highest cross-correlations (due to a high mutual overlap) and that cross-correlations increase with HR. They also suggest that older individuals exhibit slightly lower cross-correlations and fewer segments with high HRs. The plot of bin count versus mean HR and BPM is particularly notable, as it indicates a decrease in the number of beats in each bin with increasing age. This presence of outliers may account for the relatively high cross-correlation observed at ages above 70 at lag $\tau=1$, as well as the negative cross-correlations seen at other lags.

\subsection{Limitations}

There are a few limitations in our study. First, we rely on a single dataset of long-term ECG recordings. Although the dataset includes a relatively large number of subjects (>200), cross-validation with additional data would enhance the robustness of our findings. Secondly, the length of the recordings means that the data includes various individual activities (rest, exercise, sleep, etc.), making the dataset highly heterogeneous. These activities also vary with age, complicating the analysis of age effects, even though heart rate has been considered in the latter part of the study.

The ECG recordings also contain artifacts that are not fully addressed by our filtering procedure. In particular, there are uncertainties in the QT intervals, especially regarding T-wave detection. Furthermore, the detrending procedure influences the results, as shown in Fig.~\ref{fig:CC_sex_and_gamma_compare}(a), and our choice of a smoothing parameter $\gamma=500$ reflects a tradeoff based on qualitative reasoning.

Lastly, this study does not fully account for potential confounding factors when analyzing cross-correlations across different sexes, ages, and heart rates. However, given the relatively large number of subjects and the balanced grouping, we opted not to conduct a comprehensive statistical analysis of confounding factors.
\FloatBarrier
% Discussion
\label{ch:conclusions}
\section{Conclusions}

We have investigated the dynamical cross-correlations between RR and QT intervals in long-term ECG recordings of over 200 healthy subjects. Our primary method was time-lagged cross-correlation, which allowed us to thoroughly examine cross-correlations at various lags (time shifts), considering the relationships between the present QT interval and both preceding and succeeding RR intervals. During preprocessing, we employed the smoothness priors method to detrend the data, focusing on reducing low- or high-frequency components. The results were consistent across different levels of detrending, and for the main analysis, we primarily removed long-term trends from the RR/QT data.

Our findings indicate that cross-correlation between RR and QT intervals is highest when these intervals have the most overlap, which is expected. Notably, the second-highest cross-correlation occurs between the present QT interval and the preceding RR interval. More broadly, the cross-correlations are skewed towards negative lags, where RR intervals precede QT intervals. These observations align with previous studies, such as those examining transfer entropy between RR and QT intervals.

We observed that sex did not significantly affect RR-QT cross-correlations, with results being nearly identical between men and women. However, age had a notable impact, with a consistent decrease in cross-correlation across all lags as age increased. This finding aligns with prior research on heart rate and QT variability across different age groups. Additionally, our analysis of cross-correlation as a function of heart rate revealed a slightly positive relationship at all lags, with cross-correlation decreasing as the lag $\tau$ increased.

For future studies, we suggest the following directions. First, while this study focused on the cross-correlation between RR and QT intervals, other ECG intervals, such as ST, PR, and QRS, could also be analyzed using the methods outlined here. Exploring cross-correlations between these intervals could provide a broader view of the relationships within heart beat time series. Second, this study exclusively involved data from healthy subjects. Future research could investigate RR-QT cross-correlations in individuals with diagnosed health conditions and compare these patterns between healthy and non-healthy subjects, such as those with congestive heart failure or long-QT syndrome. It would also be valuable to examine these effects at the cellular level by analyzing the cross-correlations between beat-to-beat intervals and field potential durations in cardiomyocytes.
% Data availability
\label{ch:data_availability}
\section{Data availability}

The E-HOL-03-0202-003 dataset analysed during the current study is available at the Telemetric and Holter ECG Warehouse repository, \url{http://thew-project.org/Database/E-HOL-03-0202-003.html}.

%\bibliography{tex/references}

\begin{thebibliography}{10}
\urlstyle{rm}
\expandafter\ifx\csname url\endcsname\relax
  \def\url#1{\texttt{#1}}\fi
\expandafter\ifx\csname urlprefix\endcsname\relax\def\urlprefix{URL }\fi
\expandafter\ifx\csname doiprefix\endcsname\relax\def\doiprefix{DOI: }\fi
\providecommand{\bibinfo}[2]{#2}
\providecommand{\eprint}[2][]{\url{#2}}

\bibitem{LuoShen2004Acoc}
\bibinfo{author}{Luo, S.}, \bibinfo{author}{Michler, K.}, \bibinfo{author}{Johnston, P.} \& \bibinfo{author}{Macfarlane, P.~W.}
\newblock \bibinfo{journal}{\bibinfo{title}{A comparison of commonly used qt correction formulae: The effect of heart rate on the qtc of normal ecgs}}.
\newblock {\emph{\JournalTitle{Journal of electrocardiology}}} \textbf{\bibinfo{volume}{37}}, \bibinfo{pages}{81--90} (\bibinfo{year}{2004}).

\bibitem{Batchvarov2002}
\bibinfo{author}{Batchvarov, V.~N.} \emph{et~al.}
\newblock \bibinfo{journal}{\bibinfo{title}{Qt-rr relationship in healthy subjects exhibits substantial intersubject variability and high intrasubject stability}}.
\newblock {\emph{\JournalTitle{American Journal of Physiology-Heart and Circulatory Physiology}}} \textbf{\bibinfo{volume}{282}}, \bibinfo{pages}{H2356–H2363}, \doiprefix\url{10.1152/ajpheart.00860.2001} (\bibinfo{year}{2002}).

\bibitem{Potapov}
\bibinfo{author}{Potapov, I.} \emph{et~al.}
\newblock \bibinfo{journal}{\bibinfo{title}{Information transfer in qt-rr dynamics: Application to qt-correction}}.
\newblock {\emph{\JournalTitle{Scientific reports}}} \textbf{\bibinfo{volume}{8}}, \bibinfo{pages}{14992--9} (\bibinfo{year}{2018}).

\bibitem{AccuQT}
\bibinfo{author}{Räsänen, E.} \emph{et~al.}
\newblock \bibinfo{journal}{\bibinfo{title}{Accurate qt correction method from transfer entropy}}.
\newblock {\emph{\JournalTitle{Cardiovascular digital health journal}}} \textbf{\bibinfo{volume}{4}}, \bibinfo{pages}{1--8} (\bibinfo{year}{2023}).

\bibitem{TseGary2017Emol}
\bibinfo{author}{Tse, G.}, \bibinfo{author}{Chan, Y. W.~F.}, \bibinfo{author}{Keung, W.} \& \bibinfo{author}{Yan, B.~P.}
\newblock \bibinfo{journal}{\bibinfo{title}{Electrophysiological mechanisms of long and short qt syndromes}}.
\newblock {\emph{\JournalTitle{International journal of cardiology. Heart \& vasculature}}} \textbf{\bibinfo{volume}{14}}, \bibinfo{pages}{8--13} (\bibinfo{year}{2017}).

\bibitem{Tavernier1997}
\bibinfo{author}{Tavernier, R.}, \bibinfo{author}{Jordaens, L.}, \bibinfo{author}{Haerynck, F.}, \bibinfo{author}{Derycke, E.} \& \bibinfo{author}{Clement, D.~L.}
\newblock \bibinfo{journal}{\bibinfo{title}{Changes in the qt interval and its adaptation to rate, assessed with continuous electrocardiographic recordings in patients with ventricular fibrillation, as compared to normal individuals without arrhythmias}}.
\newblock {\emph{\JournalTitle{European Heart Journal}}} \textbf{\bibinfo{volume}{18}}, \bibinfo{pages}{994–999}, \doiprefix\url{10.1093/oxfordjournals.eurheartj.a015389} (\bibinfo{year}{1997}).

\bibitem{Gravel2017}
\bibinfo{author}{Gravel, H.}, \bibinfo{author}{Jacquemet, V.}, \bibinfo{author}{Dahdah, N.} \& \bibinfo{author}{Curnier, D.}
\newblock \bibinfo{journal}{\bibinfo{title}{Clinical applications of qt/rr hysteresis assessment: A systematic review}}.
\newblock {\emph{\JournalTitle{Annals of Noninvasive Electrocardiology}}} \textbf{\bibinfo{volume}{23}}, \doiprefix\url{10.1111/anec.12514} (\bibinfo{year}{2017}).

\bibitem{Malik2008}
\bibinfo{author}{Malik, M.}, \bibinfo{author}{Hnatkova, K.}, \bibinfo{author}{Novotny, T.} \& \bibinfo{author}{Schmidt, G.}
\newblock \bibinfo{journal}{\bibinfo{title}{Subject-specific profiles of qt/rr hysteresis}}.
\newblock {\emph{\JournalTitle{American Journal of Physiology-Heart and Circulatory Physiology}}} \textbf{\bibinfo{volume}{295}}, \bibinfo{pages}{H2356–H2363}, \doiprefix\url{10.1152/ajpheart.00625.2008} (\bibinfo{year}{2008}).

\bibitem{Perez2023-nm}
\bibinfo{author}{P{\'e}rez, C.} \emph{et~al.}
\newblock \bibinfo{journal}{\bibinfo{title}{The role of $\beta$-adrenergic stimulation in {QT} interval adaptation to heart rate during stress test}}.
\newblock {\emph{\JournalTitle{PLoS One}}} \textbf{\bibinfo{volume}{18}}, \bibinfo{pages}{e0280901} (\bibinfo{year}{2023}).

\bibitem{Ferreira2023}
\bibinfo{author}{Ferreira, M.} \& \bibinfo{author}{Rodriguez, M.}
\newblock \bibinfo{title}{Exploring time series correlation} (\bibinfo{year}{2023}).

\bibitem{Jacquemet2017}
\bibinfo{author}{Jacquemet, V.}, \bibinfo{author}{Gravel, H.}, \bibinfo{author}{Curnier, D.} \& \bibinfo{author}{Vinet, A.}
\newblock \bibinfo{journal}{\bibinfo{title}{Theoretical and experimental comparison of lag-based and time-based exponential moving average models of qt hysteresis}}.
\newblock {\emph{\JournalTitle{Physiological Measurement}}} \textbf{\bibinfo{volume}{38}}, \bibinfo{pages}{1885–1905}, \doiprefix\url{10.1088/1361-6579/aa8b59} (\bibinfo{year}{2017}).

\bibitem{Rochester}
\bibinfo{author}{\relax {University of Rochester Medical Center}}.
\newblock \bibinfo{title}{Identification : E-hol-03-0202-003} (\bibinfo{year}{2012}).

\bibitem{CoudercTHEW2010}
\bibinfo{author}{Couderc, J.-P.}
\newblock \bibinfo{journal}{\bibinfo{title}{A unique digital electrocardiographic repository for the development of quantitative electrocardiography and cardiac safety: the telemetric and holter ecg warehouse (thew)}}.
\newblock {\emph{\JournalTitle{Journal of Electrocardiology}}} \textbf{\bibinfo{volume}{43}}, \bibinfo{pages}{595--600}, \doiprefix\url{https://doi.org/10.1016/j.jelectrocard.2010.07.015} (\bibinfo{year}{2010}).

\bibitem{CoudercTHEW2012}
\bibinfo{author}{Couderc, J.-P.}
\newblock \bibinfo{journal}{\bibinfo{title}{The telemetric and holter ecg warehouse (thew): The first three years of development and research}}.
\newblock {\emph{\JournalTitle{Journal of electrocardiology}}} \textbf{\bibinfo{volume}{45}}, \bibinfo{pages}{677--683} (\bibinfo{year}{2012}).

\bibitem{scikit-learn}
\bibinfo{author}{Pedregosa, F.} \emph{et~al.}
\newblock \bibinfo{journal}{\bibinfo{title}{Scikit-learn: Machine learning in {P}ython}}.
\newblock {\emph{\JournalTitle{Journal of Machine Learning Research}}} \textbf{\bibinfo{volume}{12}}, \bibinfo{pages}{2825--2830} (\bibinfo{year}{2011}).

\bibitem{Genshiro}
\bibinfo{author}{Gersch, W.} \& \bibinfo{author}{Kitagawa, G.}
\newblock \emph{\bibinfo{title}{Smoothness Priors Analysis of Time Series}}, vol. \bibinfo{volume}{116} of \emph{\bibinfo{series}{Lecture Notes in Statistics}} (\bibinfo{publisher}{Springer}, \bibinfo{year}{2012}).

\bibitem{TarvainenSmoothnessPriors}
\bibinfo{author}{Tarvainen, M.}, \bibinfo{author}{Ranta-aho, P.} \& \bibinfo{author}{Karjalainen, P.}
\newblock \bibinfo{journal}{\bibinfo{title}{An advanced detrending method with application to hrv analysis}}.
\newblock {\emph{\JournalTitle{IEEE transactions on biomedical engineering}}} \textbf{\bibinfo{volume}{49}}, \bibinfo{pages}{172--175} (\bibinfo{year}{2002}).

\bibitem{Penny}
\bibinfo{author}{Penny, W.~D.}
\newblock \bibinfo{title}{Signal processing course lecture notes: Chapter 7. multiple time series} (\bibinfo{year}{2009}).

\bibitem{MarkowskiConditionsForVariance}
\bibinfo{author}{Markowski, C.~A.} \& \bibinfo{author}{Markowski, E.~P.}
\newblock \bibinfo{journal}{\bibinfo{title}{Conditions for the effectiveness of a preliminary test of variance}}.
\newblock {\emph{\JournalTitle{The American Statistician}}} \textbf{\bibinfo{volume}{44}}, \bibinfo{pages}{322--326}, \doiprefix\url{10.1080/00031305.1990.10475752} (\bibinfo{year}{1990}).
\newblock \eprint{https://www.tandfonline.com/doi/pdf/10.1080/00031305.1990.10475752}.

\bibitem{Rautaharju}
\bibinfo{author}{Rautaharju, P.~M.}
\newblock \emph{\bibinfo{title}{The Female Electrocardiogram: Special Repolarization Features, Gender Differences, and the Risk of Adverse Cardiac Events}} (\bibinfo{publisher}{Springer International Publishing AG}, \bibinfo{address}{Cham}, \bibinfo{year}{2015}), \bibinfo{edition}{2015} edn.

\bibitem{Baumert2013}
\bibinfo{author}{Baumert, M.}, \bibinfo{author}{Czippelova, B.}, \bibinfo{author}{Porta, A.} \& \bibinfo{author}{Javorka, M.}
\newblock \bibinfo{journal}{\bibinfo{title}{Decoupling of qt interval variability from heart rate variability with ageing}}.
\newblock {\emph{\JournalTitle{Physiological Measurement}}} \textbf{\bibinfo{volume}{34}}, \bibinfo{pages}{1435–1448}, \doiprefix\url{10.1088/0967-3334/34/11/1435} (\bibinfo{year}{2013}).

\bibitem{Silver}
\bibinfo{author}{Silver, N.~C.} \& \bibinfo{author}{Dunlap, W.~P.}
\newblock \bibinfo{journal}{\bibinfo{title}{Averaging correlation coefficients: Should fisher's z transformation be used?}}
\newblock {\emph{\JournalTitle{Journal of applied psychology}}} \textbf{\bibinfo{volume}{72}}, \bibinfo{pages}{146--148} (\bibinfo{year}{1987}).

\bibitem{ToivonenArticle}
\bibinfo{author}{Toivonen, E.} \& \bibinfo{author}{Räsänen, E.}
\newblock \bibinfo{journal}{\bibinfo{title}{Time-series analysis approach to the characteristics and correlations of wastewater variables measured in paper industry}}.
\newblock {\emph{\JournalTitle{Journal of Water Process Engineering}}} \textbf{\bibinfo{volume}{61}}, \bibinfo{pages}{105231}, \doiprefix\url{https://doi.org/10.1016/j.jwpe.2024.105231} (\bibinfo{year}{2024}).

\end{thebibliography}

\section*{Acknowledgements}

We are grateful to Matti Molkkari for methodological support, including the implementation of smoothness priors detrending based on Ref.~\cite{TarvainenSmoothnessPriors}. We also thank Esko Toivonen for assistance with the time-lagged cross-correlation method.

The authors acknowledge CSC – IT Center for Science, Finland, Business Finland / Research to Business (R2B) MoniCardi Project (Grant No. 1426/31/2022), The Kalle Kaihari Heart Research Fund of The University of Tampere Foundation, The Finnish Foundation for Cardiovascular Research (Grant No. 230078) and Elli and Elvi Oksanen Fund of the Finnish Cultural Foundation Pirkanmaa Regional Fund (Grant No. 50231659).

\section*{Author contributions statement}

J.K. performed the numerical simulations and analysis, and wrote the first draft of the manuscript. J.K-P. contributed to the numerical simulations and analysis. T.P. contributed to the analysis of heart-rate dependence. E.R. designed and supervised the study. All authors contributed to the writing and review of the manuscript.

\section*{Additional information}

\textbf{Competing interests}: E.R. and T.P. are shareholders of MoniCardi Ltd., a medical software company focusing on ECG analysis. The company has no interests related to this work.
%The corresponding author is responsible for submitting a \href{http://www.nature.com/srep/policies/index.html#competing}{competing interests statement} on behalf of all authors of the paper. This statement must be included in the submitted article file.
\end{document}